\documentclass{aa}
\usepackage{graphicx}
\usepackage{longtable}
\usepackage{latexsym}
\usepackage{amssymb}
\usepackage{lscape}
\input{psfig.txt}
\begin{document}
\title{H$\alpha$ surface photometry of galaxies in nearby clusters.\\ V: the survey completion.
\thanks{Based on observations taken at the observatory of San Pedro Martir (Baja California, Mexico) of the
Observatorio Astron\'omico Nacional (OAN), Universitad Autonoma de Mexico (UNAM)}
}

\author{G. Gavazzi \inst{1}
\and A. Boselli \inst{2}
\and L. Cortese \inst{1,2}
\and I. Arosio \inst{1}
\and A. Gallazzi \inst{3}
\and P. Pedotti \inst{1}
\and L. Carrasco \inst{4}
}

\authorrunning{G. Gavazzi}
\titlerunning{H$\alpha$ surface photometry of galaxies in nearby clusters}
   
\offprints{G. Gavazzi}
   
\institute{
Universit\`a degli Studi di Milano-Bicocca, Piazza delle scienze 3, 20126 Milano, Italy\\
\email {giuseppe.gavazzi@mib.infn.it}
\and
Laboratoire d'Astrophysique de Marseille, BP8, Traverse du Siphon, F-13376 Marseille
Cedex 12, France
\and Max-Planck-Institut f\"ur Astrophysik, Karl-Schwarzschild-Str. 1,
D-85748 Garching bei M\"unchen, Germany
\and
Instituto Nacional de Astrof\' \i sica, Optica y Electr\'onica,
Apartado Postal 51. C.P. 72000 Puebla, Pue., M\'exico
}
\date{}

\abstract{
We present the H$\alpha$ imaging observations of 273 late-type galaxies in the nearby rich galaxy clusters Virgo, A1367, Coma,
Cancer, Hercules and in the Great Wall, carried out primarily
with the 2.1m telescope of the San Pedro Martir Observatory (SPM) and 
with the ESO/3.6m telescope. We derived the H$\alpha$+[NII] fluxes and equivalent widths.
The H$\alpha$  survey reached completion for an optically selected sample of nearby galaxies in and outside
rich clusters. Taking advantage of the completeness of the data set, the dependence of 
H$\alpha$ properties on the Hubble type was determined for late-type galaxies in the Virgo cluster.
Differences in the gaseous content partly account for the 
large scatter of the H$\alpha$ E.W. within each Hubble-type class. 
We studied the radial distributions of the H$\alpha$ E.W. around Coma+A1367 and the Virgo clusters
in two luminosity bins. Luminous galaxies show a decrease in their average H$\alpha$ E.W.
in the inner $\sim$ 1 virial radius, while low-luminosity galaxies do not show this trend.
\keywords{Galaxies: Galaxies: photometry; Galaxies: clusters: individual: Virgo, Coma, A1367, Cancer}
}

\maketitle

%
%________________________________________________________________

\section{Introduction}

The H$\alpha$ emission arising from HII regions is the most direct indicator of the
current ($<$ 4 10$^6$ yrs), massive ($>$ 8 \rm M$\odot$) star-formation activity in galaxies (Kennicutt 1998).
Extensive H$\alpha$ imaging surveys of galaxies are being undertaken by several investigators,
among others:  Almoznino \& Brosch (1998);  
Moss \& Whittle (2000); Sakai et al. (2001); Koopmann et al. (2001); James et al. (2004).
On our part  we undertook an H$\alpha$ observing campaign of galaxies in nearby clusters and superclusters
that originated in 1990 using the aperture photometer at the KPNO 1.3m telescope (Gavazzi et al. 1991).
As soon as imaging capabilities were offered by panoramic detectors, 
it continued with the use of various other facilities. 
A generous time allocation was obtained at the San Pedro Martir 2.1m telescope, 
allowing us to observe galaxies in the Cancer cluster 
and in the two clusters (Coma and A1367) in the Great Wall (Gavazzi et al. 1998). Taking advantage of the large field 
of view offered by the WFC at the INT, a 1 sq deg. region of each of these clusters was 
fully imaged (Iglesias-Paramo et al. 2002).
Later on we focused on the Virgo cluster, again making extensive use of the SPM telescope 
(Gavazzi et al. 2002a, Paper I of this series; 2002b, Paper IV), of 
the 1.2m OHP and Calar Alto telescopes (Boselli \& Gavazzi 2002, Paper II) and of
the 2m INT and NOT telescopes (Boselli et al. 2002, Paper III).

 The present set of 273 H$\alpha$ observations of fainter targets, as described in Sect. \ref{newdata},
represents the near completion of the H$\alpha$ survey of late-type members of the Virgo
cluster down to 18.0 mag and of members of the Coma 
supercluster and of the Cancer cluster down to the limiting magnitude of 15.7.
All products of the present H$\alpha$ survey, including images, are available worldwide from the site GOLDmine
("http://goldmine.mib.infn.it/") (Gavazzi et al. 2003).

 A forthcoming paper of this series (Paper VI, in preparation) will analyze the full H$\alpha$ data-set in conjunction with
other parameters from GOLDmine, e.g. the HI properties, with the aim of deriving the radial distribution of the
star-forming regions as a function of the gaseous properties of galaxies in and outside rich clusters.

\section{The new data}
\label{newdata}
\subsection{Selection criteria}

Observations of 273 galaxies were obtained in the Virgo and Cancer clusters 
in the Coma/A1367 and  Hercules superclusters according to the following selection criteria:
\begin{itemize}
\item{their Hubble type is Sa or later (Sa-Im-BCD).}
\item{they have a measured redshift.}
\item{{\bf Virgo cluster}: they are brighter than $m_{pg}= 18.0$ mag. in the Virgo Cluster 
Catalogue (VCC) of Binggeli et al. (1985); they are classified as members of cluster A or B, as possible members or belonging to the
W, W', M clouds or to the southern extension (Binggeli et al. 1985, 1993; Gavazzi et al. 1999a); their redshift is $V<3000$  $\rm km ~s^{-1}$;
127 Virgo galaxies matching these selection criteria were observed in this work.}
\item{{\bf Coma/A1367 supercluster}: they are brighter than $m_{pg}= 15.7$ mag. in the
CGCG (Zwicky et al. 1961-68); 
they are members of the Coma (A1656) or A1367 clusters (C) to groups (G), multiplets (M) or they
are isolated members (I) of the Great Wall ($4000<V<10000$ $\rm km ~s^{-1}$) according to Gavazzi et al. (1999b); 
104 galaxies matching these criteria are reported.}
\item{{\bf Cancer cluster}:  they are brighter than $m_{pg}= 15.7$ mag. in the
CGCG (Zwicky et al. 1961-68) and are members of any of the subunits constituting the Cancer cluster
according to Bothun et al. (1983); 
33 galaxies matching these criteria are reported. }
\item{{\bf Hercules supercluster}: they are brighter than $m_{pg}= 15.7$ mag. in the
CGCG (Zwicky et al. 1961-68) and belong to either A2147 or A2151 or A2197 or A2199; 
9 galaxies matching these criteria were observed as fillers.}
\end{itemize}

The 273 target galaxies are listed in Table \ref{tabtarget} as follows:

\begin{itemize}
\item{Column 1: VCC/CGCG designation, from Binggeli et al. (1985) for Virgo,
or from Zwicky et al. (1961-68) for the Coma supercluster, Cancer and Hercules.}
\item{Column 2: NGC/IC name.}
\item{Column 3: UGC name (Nilson 1973).}
\item{Columns 4 and 5: J2000 celestial coordinates from GOLDmine.}
\item{Columns 6 and 7: major and minor optical diameters (arcmin).We measured for VCC galaxies the
diameters on the du Pont plates at the faintest detectable isophote, 
as listed in the VCC. For CGCG galaxies, these are the major and minor optical diameters 
($a_{25}$, $b_{25}$) that were derived, as explained in Gavazzi \& Boselli (1996).}
\item{Column 8: heliocentric velocity (km s$^{-1}$) from the VCC or from Gavazzi et al. (1999a, 1999b) or from NED.}
\item{Column 9: membership in a cluster or supercluster, defined as in Gavazzi et al. (1999a) 
for Virgo and in Gavazzi et al. (1999b) for the Coma/A1367 supercluster.}
\item{Column 10: the distance (Mpc); we assumed 17 Mpc for Virgo cluster A and the distances
given in Gavazzi et al. (1999a) for the various 
substructures of Virgo, 96 Mpc for galaxies in Coma and
91 Mpc for A1367. Cancer galaxies were assigned to individual groups, each with a mean distance.
For galaxies not belonging to the clusters, the distance was determined
from the individual redshifts, assuming $H_0$= 75 $\rm km~s^{-1}Mpc^{-1}$.}
\item{Column 11: morphological type as given in the VCC or in Gavazzi \& Boselli (1996).}
\item{Column 12: photographic magnitude, as given in the VCC or in the CGCG.}
\item{Column 13: total B magnitude determined and corrected for Galactic (according to Burstein \& Heiles 1978) 
and for internal extinction, as described in Boselli et al. (2003).}
\end{itemize}

\subsection{Observations}

  \begin{figure}[!t]
  \psfig{figure=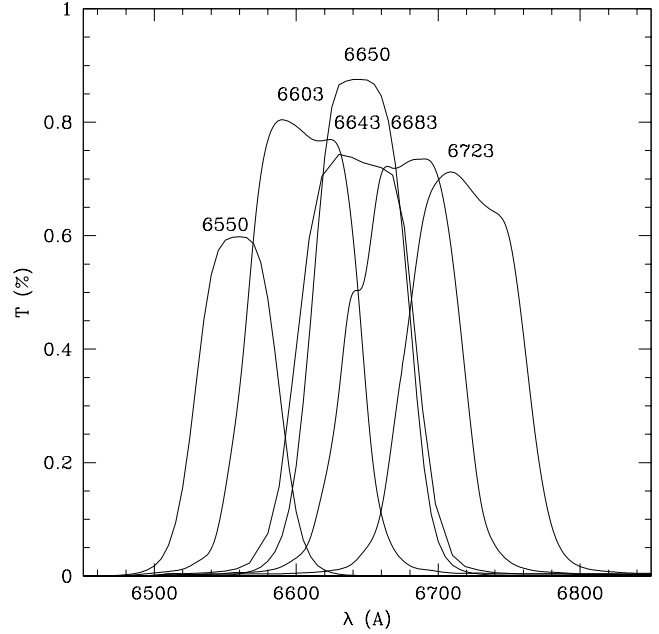,width=9cm,height=9cm}
  \caption{The transmission profiles of the adopted narrow-band filters.}
  \label{filters}
  \end{figure}
Narrow-band imaging in the H$\alpha$ emission line ($\lambda$ = 6562.8 \AA) of 273 galaxies 
listed in Table \ref{tabtarget} was obtained during 20 nights distributed in 3 runs:
2003 (Feb 26-28), 2004 (Mar 10-16) and 2005 (Apr 9-17) 
using the 2.1m telescope at San Pedro Martir Observatory (Baja California, Mexico). 
Another 18 galaxies with velocity $<$ 350 km s$^{-1}$ that could not be observed at SPM, due to the lack
of appropriate filters, were observed in 2005 (three half-nights; Mar 14-16) 
using EFOSCII attached to the ESO/3.6m telescope (proposal 074.B-0004). 

\begin{figure}[!t]
  \psfig{figure=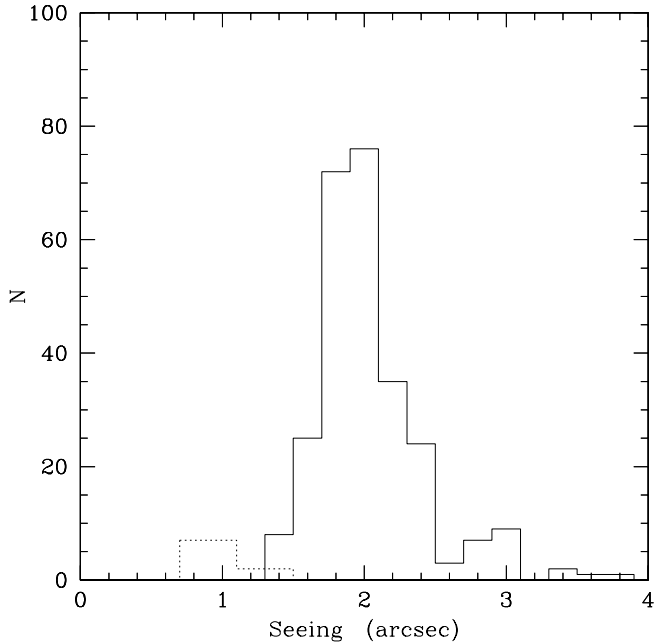,width=9cm,height=9cm}
  \caption{The seeing distribution in the ON-band frames, separately for ESO (dotted) and SPM (continuous).}
  \label{seeing}
  \end{figure}

 The (f/7.5) SPM Cassegrain focus was equipped with an 
SIT3 1024$\times$1024 pixels CCD detector with a pixel size of 0.31 arcsec.
Each galaxy was observed through a narrow band ($\sim 90$ \AA) interferometric filter centered at
the redshifted wavelength of the H$\alpha$ line, including the side [NII] lines. Four filters of similar
characteristics were used: $\lambda$ 6603, 6643, 6683, 6723  \AA, covering 
$\rm 350<V<9000 ~km~sec^{-1}$ (see Fig. \ref{filters}).
%\footnote{we could not observe galaxies in the Virgo cluster with recessional velocity $\rm V<350 ~km~sec^{-1}$}.
The stellar continuum subtraction was secured by means of shorter observations taken through
a broad-band $r$ (Gunn) filter (OFF-band frame).
The typical integration time was 15-30 min for the ON-band, sometimes split into 
shorter exposures, and 4 min for the OFF-band frame.
The observations were mostly obtained in photometric conditions, with the seeing ranging from 1.5 to 3 arcsec (see Fig. \ref{seeing}).
The observations were flux-calibrated  
using the standard stars Feige 34 and Hz44 from the catalog of Massey et al. (1988), 
observed every 2-3 hours. Periods with thin clouds were devoted to targets
that were already observed with aperture photometers providing the zero point. 

 The ESO/3.6m telescope was equipped with EFOSCII, 2048$\times$2048 pixels LORAL detector used in the $\times$ 2
rebin mode with an effective scale of 0.314 arcsec per pixel.
Two filters of $\sim 80$ \AA ~width, centered at $\lambda$ 6550 and 6650 \AA ~, provided the ON-band frames (see Fig. \ref{filters})
and the stellar continuum was taken through the $r$ (Gunn) filter.
The observations were obtained in photometric conditions, with the seeing ranging from 0.8 to 1.2 arcsec (see Fig. \ref{seeing}).
The observations were flux-calibrated  
using the standard stars LTT3864 and LTT6248 from Hamuy et al. (1994) 
observed every 2-3 hours.
Repeated measurements of standard stars obtained at either ESO or SPM in photometric conditions gave $<$ 0.05 mag differences, 
so that we assumed that the typical uncertainty ($1 \sigma$) of the 
photometric results given in this work was within 5\%. 
\subsection {Data reduction}  

The reduction of the CCD frames followed a procedure that was identical to the one 
described in previous papers (e.g. Gavazzi et al. 2002b), based on the IRAF
STSDAS \footnote{IRAF is the Image Analysis and
Reduction Facility made available to the astronomical community by the
National Optical Astronomy Observatories, which are operated by AURA,
Inc., under contract with the U.S. National Science Foundation. STSDAS
is distributed by the Space Telescope Science Institute, which is
operated by the Association of Universities for Research in Astronomy
(AURA), Inc., under NASA contract NAS 5--26555.} 
reduction packages, which is briefly summarized here. To remove the detector response, each image was
bias subtracted and divided by the median of several flat-field exposures obtained on
empty regions of the twilight sky. 
When three images in the same filter were available, a median combination of the realigned images 
allowed removal of cosmic rays. For galaxies observed in single exposures, cosmic ray
removal, as well as subtraction of contaminating objects, such as nearby stars and galaxies, was obtained by
direct inspection of the frames.
The sky background was determined in each frame in concentric object-free annuli 
around the object and then subtracted from the flat-fielded image. 
The typical uncertainty on the mean background was estimated to be 
10 \% of the rms in the individual pixels. This represents the dominant source of 
error in low S/N regions. 

\subsection{The measured H$\alpha$+[NII] parameters}

The flux from the [NII] emission lines ($\lambda$ 6548-6584~\AA) bracketing H$\alpha$ was included in the 
ON-band observations. Hereafter 
we use the simplified notation "H$\alpha$" to refer to "H$\alpha$+[NII]".
Both the flux and the E.W. of the H$\alpha$ line can be recovered from narrow ON-band
observations by subtracting the continuum contribution estimated using broad-band ($r$) images,
once these are normalized to account for the ratio of the transmissivity of the two filters.
The normalization coefficient $n$ can be derived by  
assuming that field stars (with null H$\alpha$ in emission) have identical flux
in the ON- and OFF-band frames.
By integrating the counts over diaphragms containing
the studied galaxies in the ON and OFF frames, the net counts are given by:\\
${C_{NET} = C_{ON} - n C_{OFF}.}$\\
The net flux in the H$\alpha$ line is given by
\begin{equation}
{F(H\alpha)_o = Zp \frac{C_{NET}}{T R_{ON}(6563\times(1+z))}}.
\end{equation}
\noindent
and the equivalent width by
\begin{equation}
{H\alpha ~E.W._o = \frac{\int R_{ON}(\lambda)d\lambda}{R_{ON}(6563\times(1+z))} \frac{C_{NET}}{n C_{OFF}}}.
\end{equation}
A non-negligeable ($\sim$ 10 percent) 
correction accounts for the contamination of the H$\alpha$ line emission in the 
broad band filter (OFF-band) according to (see Paper III)
\begin{equation}
{F(H\alpha) = F(H\alpha)_o (1+{\frac{\int R_{ON}(\lambda)d\lambda}{\int R_{OFF}(\lambda)d\lambda}})}
\end{equation}
and
\begin{equation}
{H\alpha E.W.=H\alpha E.W._o (1+{\frac{H\alpha E.W._o}{\int R_{OFF}}})
(1+{\frac{\int R_{ON}(\lambda)d\lambda}{\int R_{OFF}(\lambda)d\lambda}})},
\end{equation}
where
$T$ is the integration time (sec), $z$ the galaxy redshift and 
$Zp$ the ON-band zero point ($\rm erg~ cm^{-2} sec^{-1}$) 
corrected for atmospheric extinction.
$R_{ON}(\lambda)$  and $R_{OFF}(\lambda)$ are the transmissivities of the ON and OFF filters, respectively.
The errors on these parameters are determined using Eqs. 6 and 7 from Paper I.
\noindent
The results of the present observations are listed in Table \ref{tabresults} as follows:
\begin{itemize}

\item{Column 1: galaxy designation.}
\item{Column 2: year of the observation.}
\item{Column 3: Telescope}
\item{Column 4: ON-band filter}
\item{Column 5: transmissivity $R_{ON}$ of the ON-band filter at the redshifted H$\alpha$ line.}
\item{Column 6: normalization factor $n$ of the OFF-band frame.}
\item{Column 7: number of combined exposures times the integration times ($T_{ON}$) 
(in seconds) of the individual ON band exposures.}
\item{Column 8: photometric Zero point ($Zp$), such that 
$LogFlux=Zp+LogC_{NET}-LogT_{ON}$ [$\rm erg~ cm^{-2} sec^{-1}$].}
\item{Column 9: flag indicating whether the frame was taken under photometric (P) or through light cirrus (C).}
\item{Column 10: H$\alpha E.W.$ with error, in \AA ~as determined from this work.}
\item{Column 11: Logarithm of the H$\alpha$ flux, in $\rm erg~ cm^{-2} sec^{-1}$ ~as determined from this work.}
\item{Column 12 and 13 and 14: H$\alpha E.W.$ and flux from the literature, with references.} 
\end{itemize}     
All H$\alpha$ and OFF-band images of galaxies observed in this work can be obtained in the FITS format from
"http://goldmine.mib.infn.it/" (Gavazzi et al. 2003).

\subsection{Comparison with the literature}

\begin{figure}[!t]
  \psfig{figure=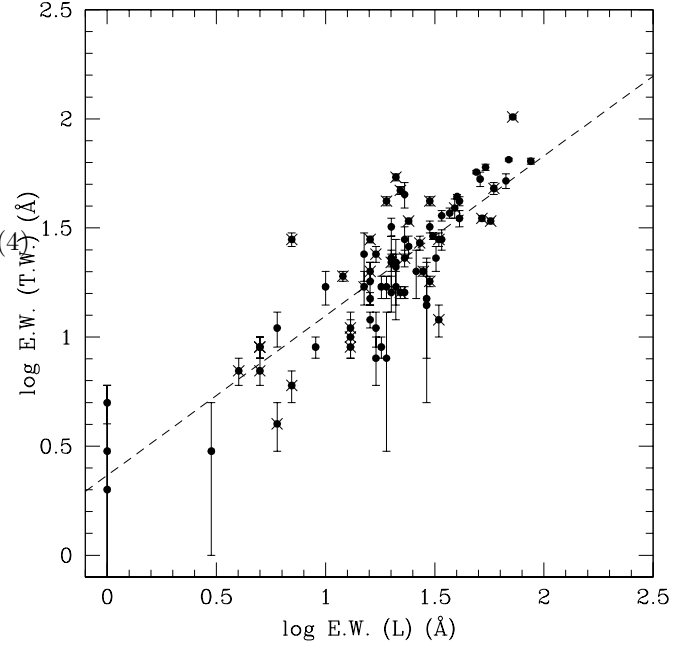,width=9cm,height=9cm}
  \caption{Comparison of the H$\alpha$ E.W. of galaxies measured in this work and those in the literature.
  The broken line represents the linear regression.}
  \label{EW}
  \end{figure}

\begin{figure}[!t]
  \psfig{figure=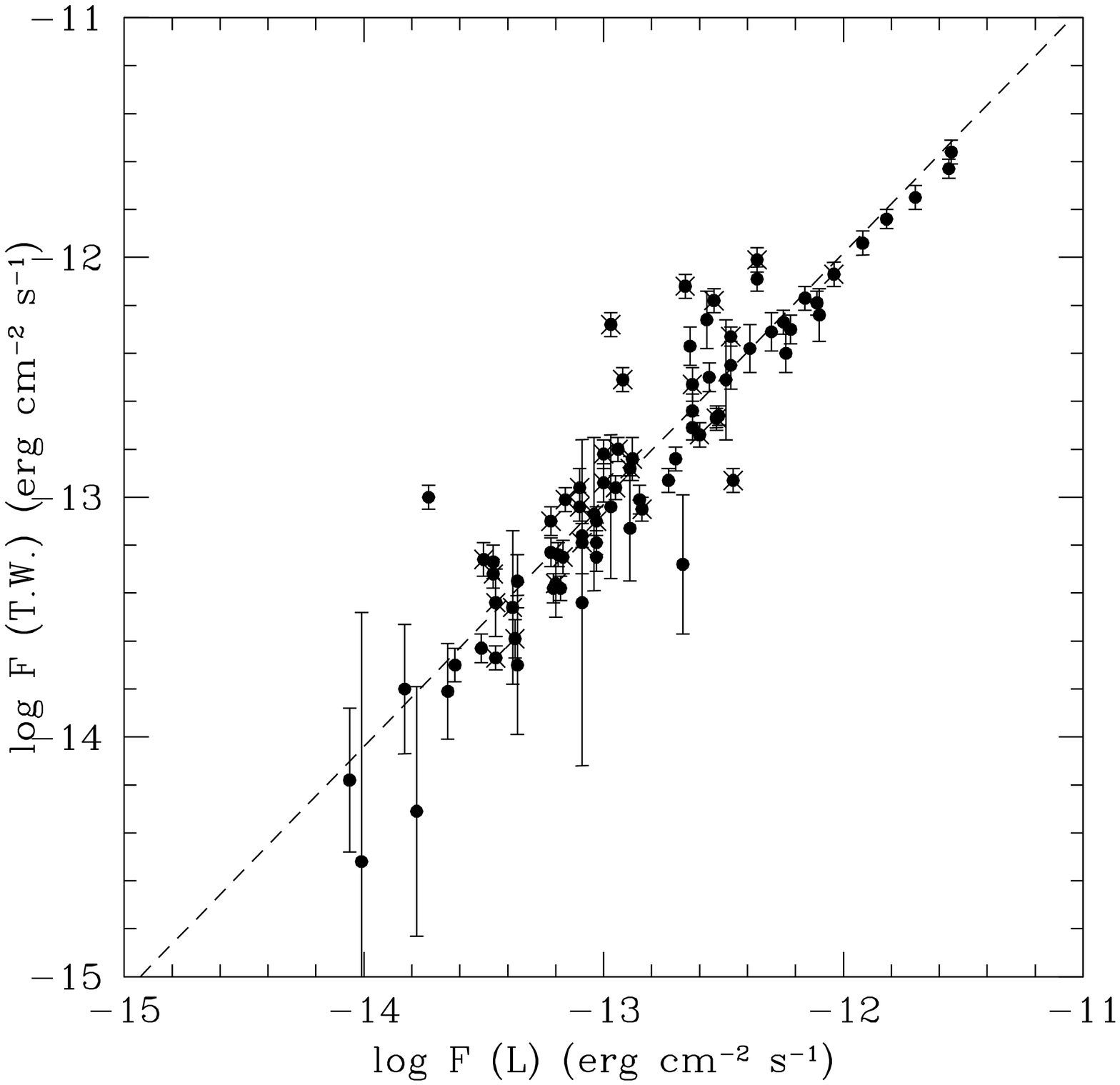,width=9cm,height=9cm}
  \caption{Comparison of the H$\alpha$ flux of galaxies measured in this work and those in the literature.
  The broken line represents the linear regression.}
  \label{flux}
  \end{figure}

Among the 56 galaxies measured in this work that have independent measurements in the literature (see Table \ref{tabresults}),
we selected those observed in photometric conditions. The two sets of measurements are in satisfactory agreement,
as illustrated in Figs. \ref{EW} and \ref{flux}.
The linear regression between our equivalent width  and flux  measurements and those found in the literature 
are:\\ 
H$\alpha E.W._{TW} = 0.84 \times H\alpha E.W._{L} +4.44$\\ 
and \\ 
$logF(H\alpha)_{TW} = 1.03 \times logF(H\alpha)_L +0.36$.\\
Literature measurements taken with aperture photometers 
(crosses in Figs. \ref{EW} and \ref{flux}) show larger discrepancies than average.


\section{Results and discussion}
\label{analysis}
  \begin{figure*}
  \psfig{figure=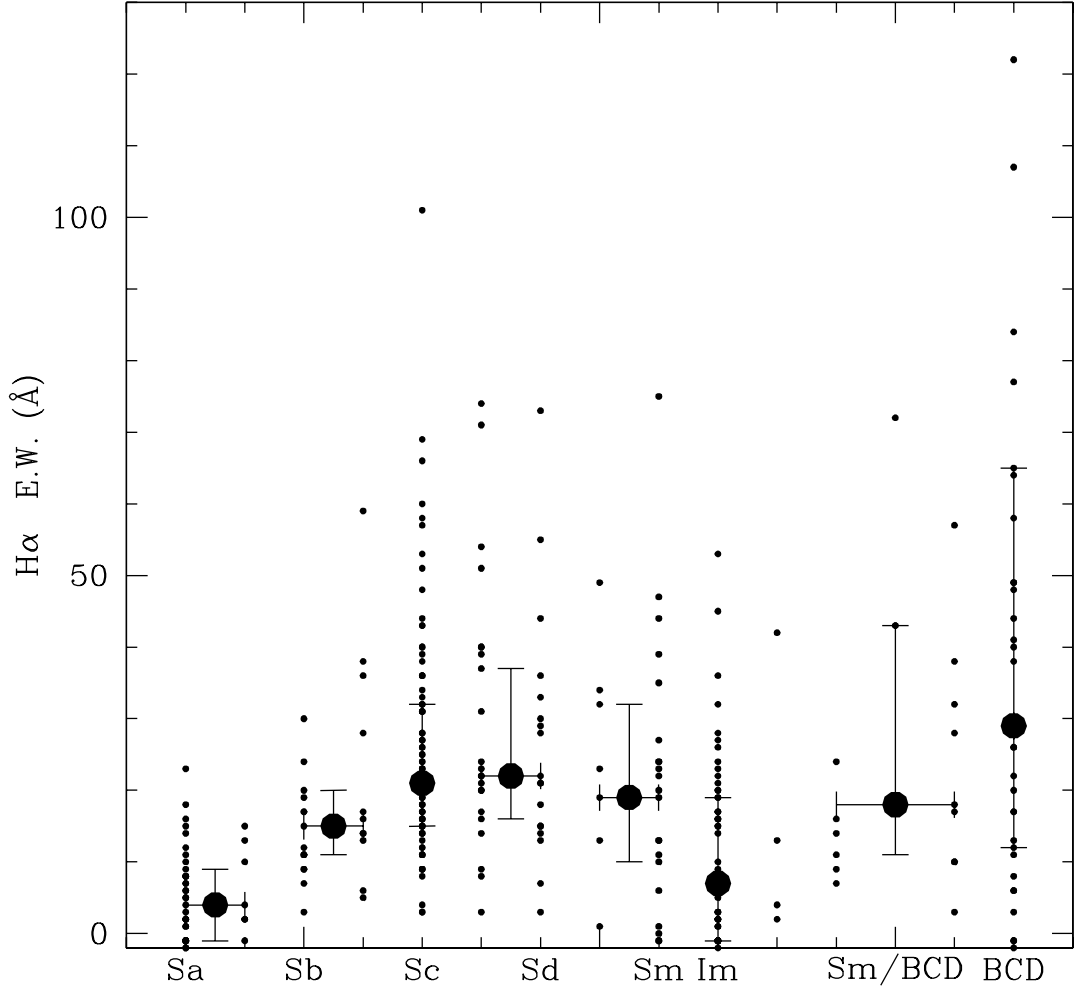,width=15cm,height=15cm}
  \caption{Distribution of the individual H$\alpha E.W.$ measurements in the 
  Virgo cluster along the Hubble sequence (small dots) and 
  of the median H$\alpha E.W.$ in bins of Hubble type. Error bars are drawn at the $25^{th}$ and $75^{th}$ percentile
  of the distribution.}
  \label{hatype}
  \end{figure*} 
    \begin{figure*}
  \psfig{figure=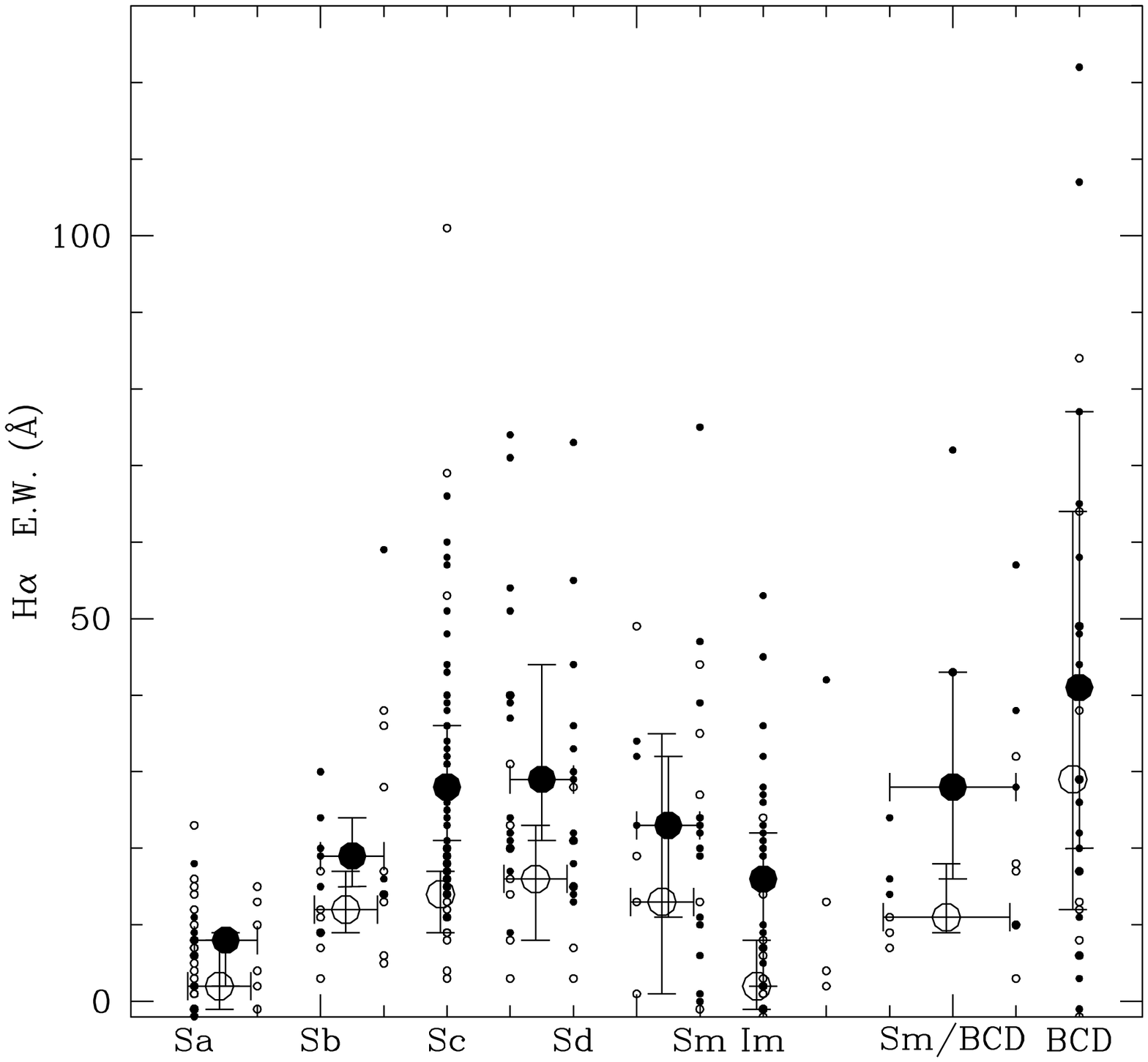,width=15cm,height=15cm}
  \caption{Same as Fig. \ref{hatype}, except that filled symbols represent 
  $Def_{HI}\leq0.4$ (unperturbed) objects and 
  open symbols $Def_{HI}>0.4$  (HI deficient) galaxies.}
  \label{hatydef}
  \end{figure*}
 Including the observations presented in this work, optically selected samples of Virgo, Cancer and Coma galaxies (excluding the Hercules
supercluster that was only observed as a filler) were nearly 
completed with H$\alpha$ imaging observations. 
Table \ref{compl}
gives the details. Among 384 Virgo late-type members within the completeness
limit of the VCC ($m_{pg}\leq18.0$), less than two dozen galaxies still need to be observed. 
In the Coma supercluster (C+G+M+I) only 33 out of 272 galaxies
with $m_{pg}\leq15.7$ still need  to be observed. This fraction nearly vanishes among both the cluster members (C)
and strictly isolated objects (I).\\  
\begin{table}
\begin{tabular}{lccc}
\multicolumn{4}{l}{\footnotesize {\bf Table 3} Completeness} \\
%\caption{completeness}\\
%\multicolumn{4}{c}{}\\
                       & members & H$\alpha$  & \%\\
\hline 
Virgo  $m_{pg}\leq18.0$             &384&   362& 94\\
Coma+A1367 (C) $m_{pg}\leq15.7$     & 70&   67 & 96\\
Coma sup. (I) $m_{pg}\leq15.7$      &118&   116& 98\\
Coma sup. (G+M+I)  $m_{pg}\leq15.7$ &212&   172& 81\\
Cancer        $m_{pg}\leq15.7$      & 35&   35 &100\\
Hercules      $m_{pg}\leq15.7$      &112&   30 & 27\\
\hline 
\end{tabular}
\label{compl}
\end{table}
\subsection{Star-formation rate, Hubble type and HI content}
  \begin{figure*}[!t]
  \psfig{figure=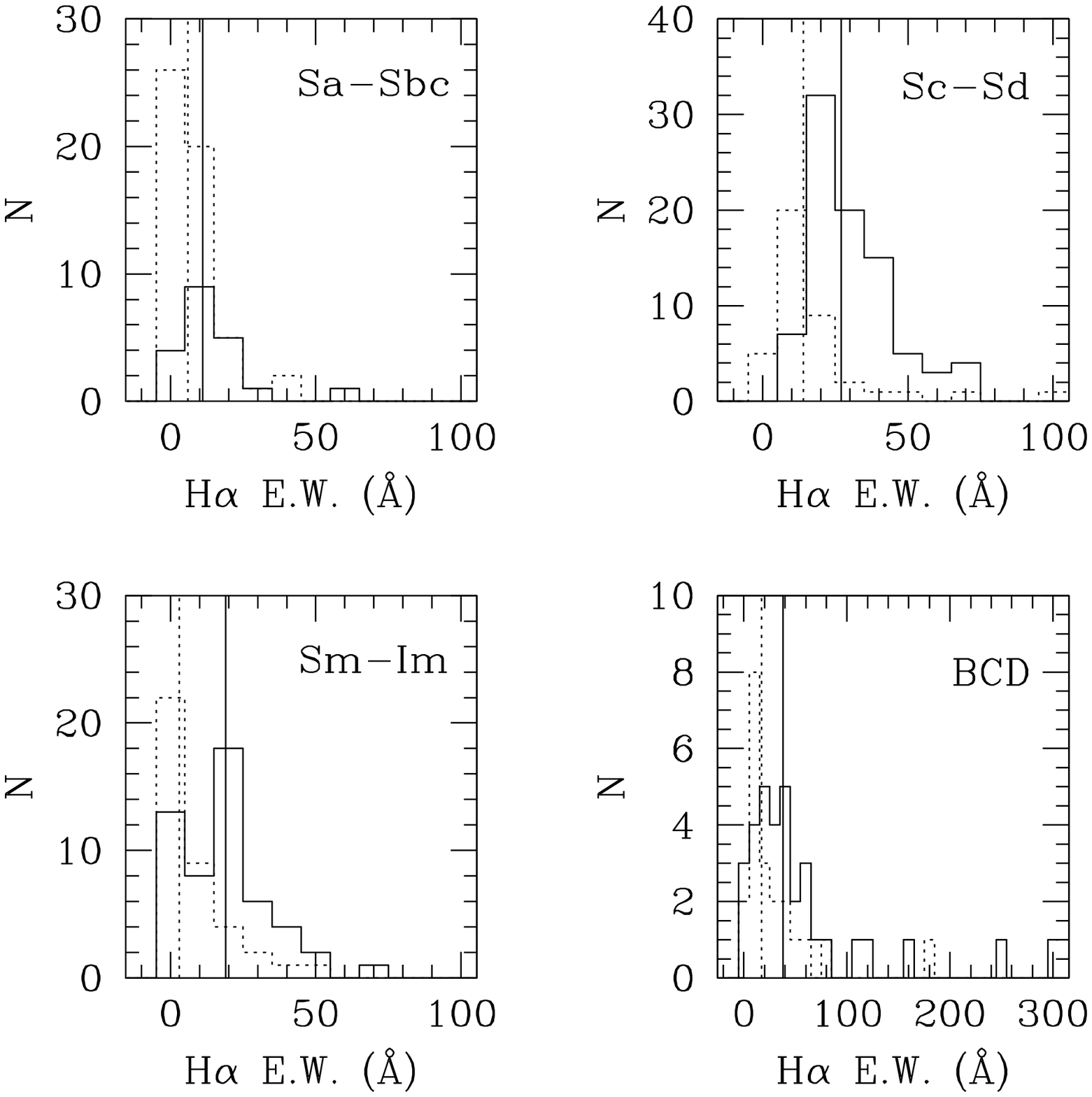,width=15cm,height=15cm}
  \caption{Distribution of H$\alpha E.W.$ in the Virgo cluster in four bins of Hubble type, divided in
  $Def_{HI}\leq0.4$  (solid histogram) and $Def_{HI}>0.4$ (dotted histogram). Vertical lines mark the median
  of each distribution.}
  \label{hahisto}
  \end{figure*}
Taking advantage of the accurate morphological classification that was achieved for the Virgo cluster, owing to
the superb plate quality of the photographic survey by Binggeli et al. (1985) 
and to the completeness of the present observations,
we wish to comment on the dependence of the star-formation 
properties of late-type galaxies in this cluster on the Hubble type.
Figure \ref{hatype} shows the distribution of the 
current massive star-formation rate per unit mass, as represented by 
the individual H$\alpha E.W.$ measurements, along the Hubble sequence.
Taking medians in bins of Hubble type, 
it appears that the star-formation rate per unit mass in giant galaxies members of the Virgo cluster
increases from Sa ($<H\alpha E.W.>=4$ ~\AA) to Sb ($<H\alpha E.W.>=15 $~\AA), Sc ($<H\alpha E.W.>=21 $~\AA) and
Scd ($<H\alpha E.W.>=22 $~\AA); then 
it decreases to a relative minimum for Im ($<H\alpha E.W.>=7 $~\AA) and 
reaches the highest value for BCDs ($<H\alpha E.W.>=29 $~\AA) (cf. similar results for 
isolated galaxies in Fig. 3 of Kennicutt 1998).
In bins of all types the dispersion of the data exceeds tens of \AA, reflecting a dependence of
 the H$\alpha E.W.$ on other parameters,
within each Hubble class.
The main dependency is the one on the galaxy global HI content, as already noted by Boselli et al. (2001)
 and Gavazzi et al. (2002b). 
Using the complete compilation of HI deficiency parameters of galaxies in the Virgo cluster by Gavazzi et al. (2005),
we show in Fig. \ref{hatydef} the H$\alpha E.W.$ vs. Hubble type diagram, separately for $Def_{HI}<0.4$ (unperturbed galaxies)
and $Def_{HI}>0.4$ (HI deficient objects), 
and in Fig. \ref{hahisto} the distribution of the H$\alpha E.W.$ 
in the Virgo cluster in four bins of Hubble type, given separately for $Def_{HI}<0.4$ 
and $Def_{HI}>0.4$. 
The figures emphasize that, within each Hubble type class, galaxies with normal HI content have H$\alpha E.W.$ 
 that is systematically higher than their HI deficient counterparts by a factor of two. 
Moreover, the figures show that the relative minimum found for Im galaxies is not due to
a particularly high HI deficiency among these objects, but is a characteristic
of all objects in this Hubble class.    

\subsection{The radial dependence of the star-formation rate in nearby clusters}

One of the hottest issues in galaxy evolution is whether the star-formation rate depends
on the environmental properties of galaxies, i.e. on the local galaxy density and/or
projected distances from galaxy density enhancements (clusters).
The 2dF (Lewis et al. 2002) and the SDSS surveys (Gomez et al. 2003, Goto 2003, 
Nichol 2004, Kauffmann et al. 2004) have clearly shown that the star-formation rate is found to be significantly quenched
when the galaxy density overcomes the characteristic value of $\sim$ 1 $h_{75}^{-2}$Mpc$^{-2}$, i.e.
between 1 to 2 virial radii from cluster centers, which is the regime where significant
gas depletion occurs (van Gorkom et al. 2004).
These works, however, are limited by the lack of a reliable  morphological classification 
for their objects (in fact based on light concentration and/or spectrophotometric indicators),
and thus they cannot disentangle the radial dependence of the star-formation
activity from the morphology-segregation effect.
Furthermore, SDSS and 2dF only include high-luminosity 
($M_r$ $\leq$ -20.5 for SDSS and $M_B$ $\leq$ -19 for 2dF) galaxies that are shallower or similar to our survey 
of Coma ($M_B$ $\leq$ -19.2) and much shallower than Virgo ($M_B$ $\leq$ -13).

\begin{figure}[t!]
  \psfig{figure=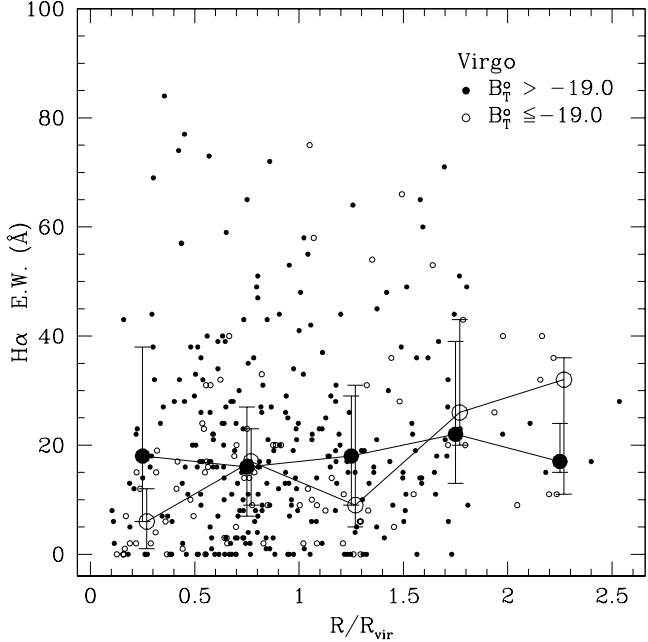,width=9cm,height=9cm}
  \caption{The clustercentric radial distribution of the individual H$\alpha E.W.$ measurements in the Virgo cluster.
   High and low (B-band) luminosity galaxies are given with open and filled dots, respectively. Median in bins of 0.5 $R/R_{Vir}$ are given.
   Error bars mark the $25^{th}$ and $75^{th}$ percentile of the distribution.}
  \label{haradvirgo}
  \end{figure}

\begin{figure}[t!]
  \psfig{figure=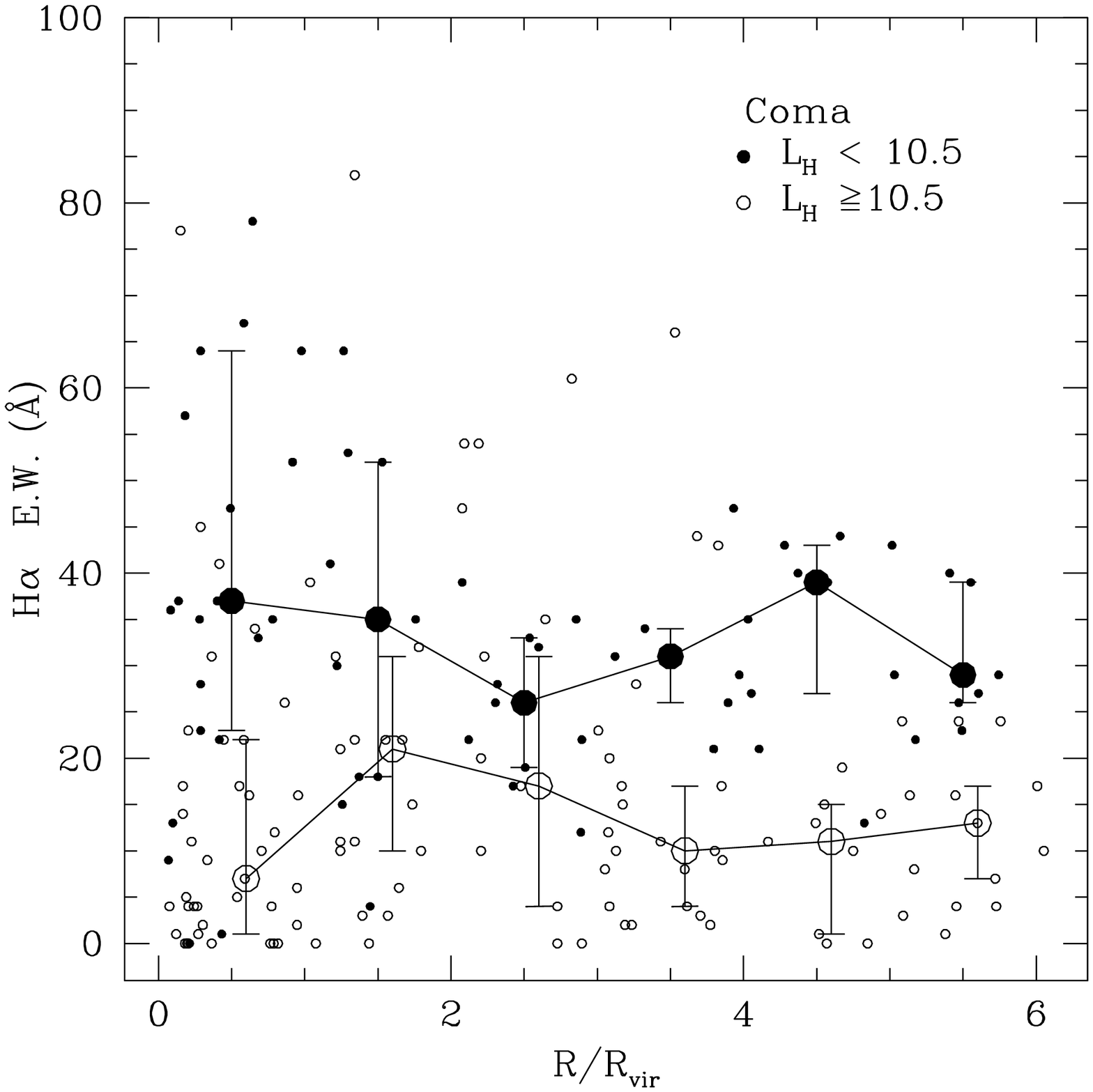,width=9cm,height=9cm}
  \caption{The clustercentric radial distribution of the individual H$\alpha E.W.$ measurements 
  in the Coma+A1367 clusters and in the Great Wall.
   High and low (H-band) luminosity galaxies are given with open and filled dots, respectively. The median in bins of 1 $R/R_{Vir}$ are given.
   Error bars mark the $25^{th}$ and $75^{th}$ percentile of the distribution.}
  \label{haradcoma}
  \end{figure}

Gavazzi et al. (2002b), analyzing the H$\alpha$ properties of spiral galaxies that were available prior to 2003,  
found indications that only high luminosity objects show the expected decrease of SFR with 
decreasing projected distance from the center of Virgo, A1367 and Coma, while low luminosity galaxies 
show an opposite trend.
 Taking advantage of the presently complete H$\alpha$ survey, we re-analyzed 
the dependence of the H$\alpha E.W.$ on 
the clustercentric radius in units of Virial radii, assuming 1.7 Mpc for Virgo and 2.2 Mpc for Coma 
(Figs.\ref{haradvirgo} and \ref{haradcoma}; Girardi et al. 1998).
Virgo galaxies are subdivided in high and low luminosity (above and below $B_T^o=-19.0$) 
according to their Blue luminosity, which is available for the majority of galaxies (see Column 13 of Table 1).
For Coma and A1367 (combined) we instead used a threshold based on the Near-IR (H band) luminosity 
available for all galaxies in the Coma supercluster, 
as it provided the best estimate of the dynamical mass within the galaxies optical extent
(Gavazzi et al. 1996). The luminosity threshold ($L_H=10.5$) adopted for Coma is on average
half a magnitude brighter than the one adopted for Virgo.  
% We confirm that the SFR per unit mass of high luminosity spirals 
% is about a factor of 2-3 lower inside 2 virial radii than at larger clustercentric
% projected distances. On the contrary there is further evidence this not being
% the case for low luminosity objects that show a constant SFR at all Virial radii.
Both distributions did not
appear to be strong functions of the clustercentric distance.
Low-luminosity galaxies have systematically higher EW than
high-luminosity ones and this difference is highest at radii below
$1.5 R_{vir}$, reflecting a possible suppression  of SFR for giant galaxies. 
The decline in SFR (as expressed by
the H$\alpha E.W.$) is clearer for giant late-type galaxies in the Virgo cluster
from 2 to 0.5 $R_{vir}$. On the contrary, we did not identify any significant
trend with clustercentric distance for the dwarf spirals.

 The different patterns of high and low luminosity could be observed by neither the 2dF nor SDSS surveys due to their
shallowness, which indicates that giant spiral galaxies within the virial radius 
were subject to gas removal long enough for their SFR to be significantly quenched. 
Dwarf galaxies instead might have had less time to develop such depletion, because rich clusters are being significantly 
replenished with small galaxies at the present cosmological epoch. 
Favoring this hypothesis, we found that low and high luminosity late-type members of these three clusters
have significantly different velocity distributions, with the distribution of the low luminosity objects  
broader and more deviating from a Gaussian than that of the high luminosity objects (see Adami et al. 1998). 
For the Virgo cluster, owing to the detailed morphological classification available from the VCC,
it is possible to divide up the 729 members with known velocity ($-1000<V<3000$ $km~sec^{-1}$) between
early and late ($>$Sa) types. Among the latter, the velocity dispersion of the 50 
brightest ($m_{pg}<12.5$) galaxies is 782 $km~sec^{-1}$. 
Fainter than $m_{pg}=12.5, 13.5, 14.5$, it becomes 1570, 1717, 1911 $km~sec^{-1}$, respectively.


\section{Summary}

The H$\alpha$ flux and equivalent width of 273 galaxies obtained with the SPM 2.1m and with the ESO/3.6m 
telescopes have been presented.
These observations conclude the long-term H$\alpha$ survey of late-type galaxies in the Virgo, 
Coma, A1367 and Cancer clusters, as well as of relatively isolated objects in the Great Wall,
spanning a wide range of morphological type, luminosity and environmental conditions.

 In the Virgo cluster in particular, all giant spirals and virtually all late-type dwarfs
were observed  down to the optical limit of $m_{pg}=18$ mag ($M_{pg}=-13$ mag).\\
From these data we conclude that:\\ 
1) the star-formation rate per unit mass in giant galaxy members of the Virgo cluster
increases from Sa to Scd, then decreases to a relative minimum for Im, finally reaching the highest peak for BCDs.\\
2)  Within each Hubble type class, galaxies with normal HI content have twice the H$\alpha E.W.$ 
of their HI-deficient counterparts. \\
3)  The SFR per unit mass of high luminosity spirals that are projected within 1 virial radius
is about a factor of two lower than at larger clustercentric
projected distances. Low luminosity objects have similar H$\alpha$ properties at all
clustercentric radii.\\
 Analysis of the full H$\alpha$ data-set, including study of the galactocentric radial distribution 
of the star-forming regions as a function of other environmental indicators (i.e. 
gaseous properties of galaxies in and outside rich clusters)
will be presented in a forthcoming paper of this series (Paper VI: The morphology of the star-forming regions, 
in preparation).

\begin{acknowledgements}
We thank the SPM TAC for the generous time allocations. 
This research has made use of the NASA/IPAC Extragalactic Database (NED) which is operated 
by the Jet Propulsion Laboratory, California Institute of Technology, under contract with the
National Aeronautics and Space Administration. We also made extensive use of the Goldmine 
(Galaxy On Line Database Milano Network) database.
A.G. thanks the Alexander von Humboldt Foundation, the Federal Ministry
of Education and Research, and the Program for Investment in the Future 
(ZIP) of the German Government for funding through a Sofja Kovalevskaja award.
\end{acknowledgements}

\newpage
%\clearpage
\onecolumn
%\scriptsize
\tiny
%
  \normalsize
 \footnotesize{References:}
\noindent
(1)  Kennicutt \& Kent 1983; 
(2)  Kennicuttet al. 1984;
(4)  Moss et al. 1988;
(6)  Gavazzi et al. 1991; 
(7)  Romanishin 1990;
(8)  Gavazzi et al. 1998; 
(14) Almoznino \&  Brosch 1998;  
(19) Moss et al.  1998; 
(22) Heller et al. 1999; 
(23) Usui et al. 1998; 
(24) Boselli \& Gavazzi 2002;
(29) Gallagher \& Hunter 1989;
(32) Koopmann et al.  2001; 
(40) James et al. 2004;
(1997, 2002, 2003) repeated measurements obtained at SPM.\\
    
Notes on individual objects:\\
VCC 710 (dS0), VCC 1617 (dS0), CGCG 14034 (SB0 in the RC3): these three galaxies were observed in spite of 
their classification as early-type because spectroscopic observations revealed emission lines. 
H$\alpha$ emission was detected in all of them suggesting that their classification should be revised. \\
CGCG 160086, 160088, 160127, 160128, 160139 re-observed due to poor quality of the previously available data.\\
CGCG 224024: affected by stray-light from bright star.\\
VCC 531: affected by stray-light from bright star.\\

\end{document}